\documentclass[12pt]{article}

\usepackage{amsmath,amsfonts}

\usepackage{slashed}

\usepackage{graphicx}
\usepackage{color}

\usepackage{hyperref} 
\hypersetup{colorlinks=true, citecolor=blue, filecolor=black, linkcolor=blue, urlcolor=blue, pdfpagemode=UseNone}

\advance \textheight by \topskip
\makeatletter
\@addtoreset{equation}{section}
 
\makeatother

\newcommand{\be}{\begin{equation}}
\newcommand{\ee}{\end{equation}}
\newcommand{\bea}{\begin{eqnarray}}
\newcommand{\eea}{\end{eqnarray}}
\newcommand{\dd}{\partial}

\def\>{\rangle}
\def\<{\langle}

\begin{document}

\title{
{\bf A classical  complex $\phi^4$ scalar field in a gauge background  }}


\author{
{\sf   N. Mohammedi} \thanks{e-mail:
noureddine.mohammedi@univ-tours.fr}$\,\,$${}$
\\
{\small ${}${\it Institut  Denis Poisson (CNRS - UMR 7013),}} \\
{\small {\it Universit\'e  de Tours,}}\\
{\small {\it Facult\'e des Sciences et Techniques,}}\\
{\small {\it Parc de Grandmont, F-37200 Tours, France.}}}
\date{}
\maketitle
\vskip-1.5cm

\vspace{2truecm}

\begin{abstract}

\noindent
We solve the equations of motion of a complex $\phi^4$ theory coupled to some given gauge field background.
The solutions are given in both cylindrical and spherical coordinates and  have finite energy.

\end{abstract}

\newpage

%

\setcounter{equation}{0}

\section{Introduction}

The equations of motion of the Abelian Higgs model have so far resisted 
all attempts to solving them analytically. These equations, on the other hand, 
carry some rich structures like vortices \cite{nielsen-olesen} and other topological entities \cite{witten,vilenkin-shellard}.
It is, therefore,  
important to find some of these classical solutions even in some very special situations.
\par
In this note, we simplify the Abelian Higgs model by 
leaving out the gauge field kinetic term.  This means that the gauge 
field is not propagating and merely furnish  a background for the dynamical
complex scalar field.
\par
The issue now lies in conveniently  choosing this gauge field background.
Our approach, in making this choice, consists in taking the complex scalar field
as a  given function and then determine the gauge field. The guiding
principles are solvability of the equations of motion  and a corresponding finite energy.
We will look for static solutions only.       
\par
This method of proceeding is inspired from other physical problems, especially in quantum mechanics:
Suppose that one is given the time-independent Schr\"odinger equation in
one dimension and asked what is the form of the potential  $V\left(x\right)$ and energy $E$
that accompany a wave function of the type
$
\Psi\left(x\right)=c\,\exp\left(-\alpha^2 x^2\right) \,\,\,\,?
$
One, of course, finds the harmonic oscillator potential  $V\left(x\right)=2\alpha^4\hbar^2\,x^2/m$ and
its ground state energy  $E=\alpha^2\hbar^2/m$. This programme is the essence of the Bijl-Jastrow ansatz for 
many-body problems \cite{bijl,jastrow}. 
\par
We know that $\phi^4$-theory, on its own,  possesses the well-studied kink solution. In this note,
we consider a gauged $\phi^4$-theory and find the gauge field background that accompany  
the kink solution. We examine the issue in cylindrical and spherical coordinates. 
In both case we give the expression of the gauge field background as well as the corresponding
energy.

\section{A complex scalar field in a gauge background}

The field theory we consider is given by the action
\bea
S
= 
\int d^4x\sqrt{-g}\left[
\left(D_\mu\phi\right)^\star \left(D^\mu\phi\right)
 -\frac{\lambda}{2}\left(\phi^\star \phi -v^2\right)^2\right]
 \,\,\,.
\label{A-H-M}
\eea
Here $A_\mu$
is  a  gauge field taken as a background in which the complex scalar field
$\phi$ evolves. 
The gauge covariant derivative is $D_\mu = \dd_\mu +ieA_\mu$. 
We take $v^2$ and $\lambda $ to be both positive . The space-time indices are raised 
and lowered with a  metric $g_{\mu\nu}$ and $g=\det{g_{\mu\nu}}$.

\par
The equation of motion for the scalar field $\phi$ is
\bea
\frac{1}{\sqrt{-g}}\,D_\mu\left(\sqrt{-g}\,D^\mu\phi\right)
+ {\lambda}\phi\left(\phi^\star \phi -v^2\right)
 &=&0\,\,\,\,\,.
\,\,\,\,\,
\label{eom2}
\eea
The corresponding  energy-momentum tensor is given by
\bea
T_{\mu\nu} &=& \frac{2}{\sqrt{-g}}\,\frac{\delta S}{\delta g^{\mu\nu}}
\nonumber \\ 
&=&  \left(D_\mu\phi\right)^\star  \left(D_\nu\phi\right)
+ \left(D_\nu\phi\right)^\star  \left(D_\mu\phi\right)
- g_{\mu\nu}\left[
 \left(D_\alpha\phi\right)^\star  \left(D^\alpha\phi\right)
 -\frac{\lambda}{2}\left(\phi^\star \phi -v^2\right)^2\right]
 \,\,\,.
\label{en-mo}
\eea

\subsection{Static solution in cylindrical coordinates}

We first use cylindrical coordinates
$x^\mu =\left(t\,,\, r\,,\, \theta\,,\, z\right)$ for which the metric is  
\bea
g_{\mu\nu}={\rm{diag}}\left(1\,,\,-1\,,\,-r^2\,,\,-1\right)\,\,\,\,\,\,.
\eea
We assume a Nielsen-Olesen \cite{nielsen-olesen} vortex  ans\"{a}tze\footnote{The vector field with index up is
 $A^\mu = \left[0\,,\,0\,,\,\frac{n}{e}\,\frac{h\left(r\right)}{r^2}\,,0\right]$.}
\bea
\phi &=& v\,f\left(r\right)\,e^{i\,n\,\theta}\,\,\,\,\,\,,
\label{ansatz-phi}
\nonumber \\
 A_\mu &=& \left[0\,,\,0\,,\,-\frac{n}{e}\,h\left(r\right)\,,0\right]
\,\,\,\,\,\,,
\label{ansatz-A}
\eea
where $n$ is an integer in $\mathbb{Z}$. 
\par
With these ans\"{a}tze, the equation of motion  becomes
\be
\left\{\frac{d^2f}{dr^2}
 -\lambda v^2f \left(f^2-1\right)\right\}
+\left\{ \frac{1}{r}\,\frac{df}{dr} -
n^2\frac{f}{r^2}\left(1-h\right)^2\right\}
 =0
\,\,\,\,\,\,.
\label{sep-eom}
\ee
The energy per unit length (along the  $z$ direction)  is  
\bea
E &=& \int_0^{2\pi} d\theta \int_0^\infty dr\,\sqrt{-g}\, T_{00}\,\,\,\,\,\,
\nonumber\\
 &=& 2\pi\,v^2\int_0^\infty dr\, r\left[
 \left(\frac{df}{dr}\right)^2 +
\frac{n^2}{r^2}\, f^2\,\left(1-h\right)^2
+\frac{\lambda}{2}\,v^2\left(f^2 -1\right)^2\right]
\,\,\,\,\,\,.
\eea
Here $T_{00}$ is read from the energy-momentum tensor (\ref{en-mo}). The equation of motion 
(\ref{sep-eom}) corresponds to the minimum of the energy functional $E$ (that is, $\delta E=0$, with $h(r)$ fixed).


Our strategy is to set each expression between the curly brackets in (\ref{sep-eom}) to
zero separately. The vanishing of the first bracket is solved by 
\bea
f\left(r\right) &=& \tanh\left(\alpha\,r\right)
\,\,\,\,\,\,,\,\,\,\,\,\, \alpha^2=\frac{\lambda v^2}{2}\,\,\,\,\,\,.
\eea
This is the well-known kink of $\phi^4$ theory (the anti-kink corresponds to 
$f\left(r\right) = - \tanh\left(\alpha\,r\right)$).
Setting the second bracket to zero  leads to
\bea
h\left(r\right) &=&  1\pm\,\frac{1}{n}\sqrt{\frac{2\alpha\,r}{\sinh\left(2\alpha\,r\right) }}
\,\,\,\,\,\,.
\eea
This determines the gauge field background.
\par
\begin{figure}[thp]
\begin{center}
\includegraphics[scale=0.4]{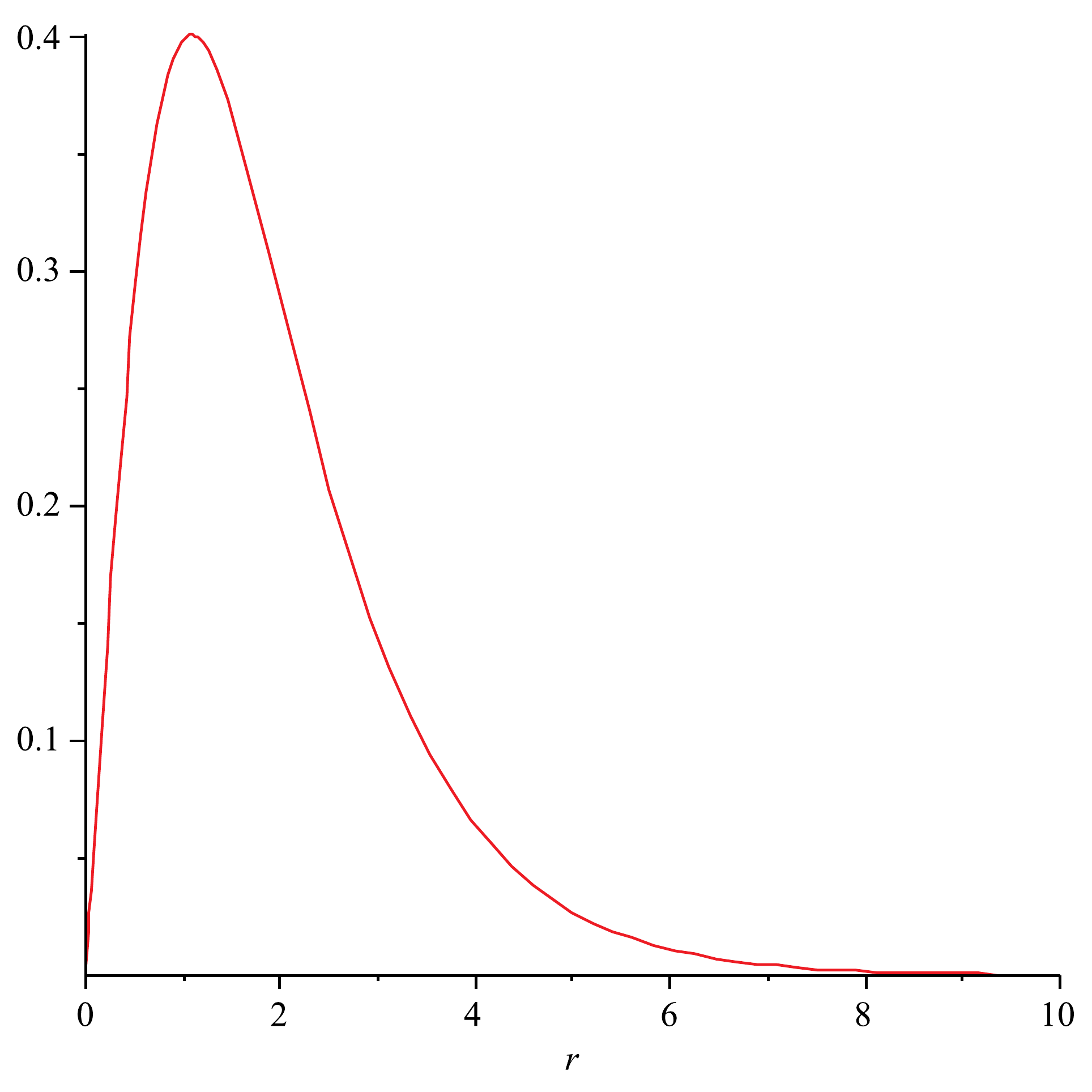}
\end{center}
\caption[]{The gauge field strength $F_{r\,\theta}$, (\ref{F-cylin}),  in cylindrical coordinates  for 
 ${e}=\pm 1$ and $\alpha=1$.}
\label{B}
\end{figure}
\begin{figure}[thp]
\begin{center}
\includegraphics[scale=0.4]{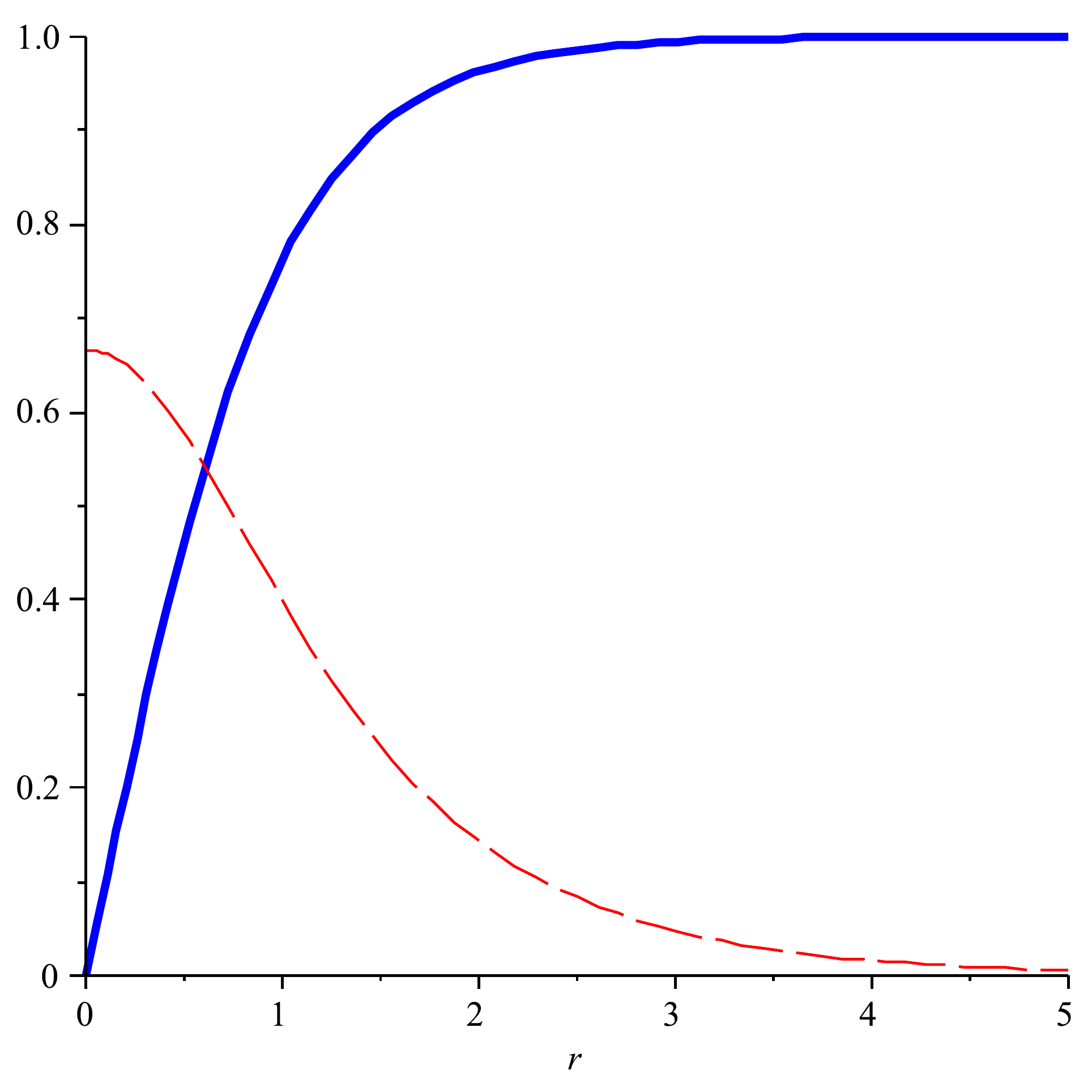}
\end{center}
\caption[]{The thick line is the kink $f\left(r\right) = \tanh\left(\alpha\,r\right)$ while the dashed line represents
the background magnetic field  $B_z$ for 
 ${e}=\pm 1$ and $\alpha=1$ in cylindrical coordinates.}
\label{f-B}
\end{figure}

The only non-vanishing component of the gauge field strength  
 $F_{\mu\nu} =  \dd_\mu A_\nu - \dd_\nu A_\mu$ is
\bea
F_{r\,\theta}=\dd_r\,A_\theta =-\frac{n}{e}\,\frac{dh}{dr}
=\mp\,\frac{1}{e}\,\frac{1}{2r}\,\sqrt{\frac {2\alpha r}  {\sinh\left(2\alpha r\right)}}
\left[1- \frac{2\alpha r}{\tanh\left(2\alpha r\right)} \right]
\,\,\,\,\,\,.
\label{F-cylin}
\eea
This is represented in figure (\ref{B}). 
The magnetic field (in a curved space-time) is defined as
\bea
B_i &=& \frac{1}{2}\sqrt{-g}\,\epsilon_{ijk}\,F^{jk}\,\,\,\,\,\,\Rightarrow\,\,\,\,\,\,
\nonumber \\
B_z &=& \frac{F_{r\,\theta} }{r} = \mp\,\frac{1}{e}\,\frac{1}{2r^2}\,\sqrt{\frac {2\alpha r}  {\sinh\left(2\alpha r\right)}}
\left[1- \frac{2\alpha r}{\tanh\left(2\alpha r\right)} \right]
\,\,\,\,\,\,,
\eea
where the flat alternating tensor is such that $\epsilon_{123}=\epsilon_{r\theta z}=+1$. For completness, we have sketshed 
in figure (\ref{f-B}) the function $f(r)$ and the magnetic field $B_z$.

\par
Using the equation of motion  (\ref{sep-eom}) and the expression of $f(r)$, the energy $E$ can be written as
\bea
E &=& 2\pi\,v^2\int_0^\infty dr\, \left[
 \frac{1}{2}  \dd_r\left(r\dd_r\,f^2\right)
-\frac{\lambda}{2}\,v^2\,r\left(f^4 -1\right)\right]
\,\,\,\,\,\,
\nonumber\\
&=&-2\pi\,v^2\alpha^2\int_0^\infty dr\,r \left[\tanh^4\left(\alpha r\right) -1 \right]
\,\,\,\,\,\,.
\eea
The total derivative term does not contribute. Performing the integral we obtain
\be
E=\frac{2\pi}{3}\,{v^2}\left[\frac{1}{2}+4\ln\left(2\right)\right] 
\,\,\,\,\,\,.
\label{E-cylinder}
\ee
We recall that  $E$  here is the energy per unit length.

\newpage

\subsection{Static solution in spherical coordinates}

The same study could be repeated in spherical coordinates 
$x^\mu =\left(t\,,\, r\,,\, \theta\,,\, \varphi\right)$ for which the metric is  
\bea
g_{\mu\nu}={\rm{diag}}\left(1\,,\,-1\,,\,-r^2\,,\,-r^2\sin^2\left(\theta\right)\right)\,\,\,\,\,\,.
\eea
The  ans\"{a}tze for the complex scalar field $\phi$ and the gauge field $A_\mu$ are
\bea
\phi &=& v\,f\left(r\right)\,e^{i\,n\,\theta}\,\,\,\,\,\,,
\label{ansatz-phi-2}
\nonumber \\
 A_\mu &=& \left[0\,,\,0\,,\,-\frac{n}{e}\,,\,-\frac{1}{e}\,\sin\left(\theta\right)\,h\left(r\right)\right]
\,\,\,\,\,\,.
\label{ansatz-A-2}
\eea

\par
The equation of motion is now
\be
\left\{\frac{d^2f}{dr^2}
 -\lambda v^2f \left(f^2-1\right)\right\}
+\left\{ \frac{2}{r}\,\frac{df}{dr} -
\frac{1}{r^2}f h^2\right\}
 =0
\,\,\,\,\,\,.
\label{sep-eom-2}
\ee
Requiring each bracket to vanish results in
\bea
f\left(r\right) &=& \tanh\left(\alpha\,r\right)
\,\,\,\,\,\,,\,\,\,\,\,\, \alpha^2=\frac{\lambda v^2}{2}\,\,\,\,\,\,,
\nonumber \\
h\left(r\right) &=& \pm\, {\sqrt{2}}\,\sqrt{\frac{2\alpha\,r}{\sinh\left(2\alpha\,r\right) }}
\,\,\,\,\,\,.
\eea

In the case at hand, the non-vanishing components of the gauge field strength are
\bea
F_{r\,\varphi} &=& \dd_r\,A_\varphi =-\frac{1}{e}\,\sin(\theta)\frac{dh}{dr}
=\mp\,\frac{1}{\sqrt{2}\,e}\,\sin(\theta)\frac{1}{r}\,\sqrt{\frac {2\alpha r}  {\sinh\left(2\alpha r\right)}}
\left[1- \frac{2\alpha r}{\tanh\left(2\alpha r\right)} \right]
\,\,\,\,\,\,,
\nonumber \\
F_{\theta\,\varphi} &=& \dd_\theta\,A_\varphi = \mp\,\frac{1}{e}\cos(\theta)\,h(r)=
\mp\,\frac{\sqrt{2}}{e}\cos(\theta)\,\sqrt{\frac {2\alpha r}  {\sinh\left(2\alpha r\right)}}
\,\,\,\,\,\,,
\eea
It can be seen that the component $F_{r\,\varphi}$ reaches  a maximum for $\theta=\pi/2$ and a very small value of $r$.
Away from this maximum  $F_{r\,\varphi}$ tends rapidly to  zero. On the other hand, $F_{\theta\,\varphi}$  has two
extrema for $\theta=0$ and  $\theta=\pi$ at a very small value of $r$ and goes to zero as $r$ increases.
\par
Using the expression $B_i = \frac{1}{2}\sqrt{-g}\,\epsilon_{ijk}\,F^{jk}$, with $\epsilon_{r\theta\varphi}=+1$, the components of the magnetic field are
\be
B_\theta= -F_{r\,\varphi} \,\,\,\,\,\,,\,\,\,\,\,\,
B_r= \frac{F_{\theta\,\varphi}}{r^2} \,\,\,\,\,\,.
\ee
These components reach their extrema around $r=0$ and  tend rapidly to zero as $r$ increases.



The corresponding energy, in the whole volume, is 
\bea
E &=&\int_0^{2\pi} d\varphi \int_0^{\pi}
 d\theta \int_0^\infty dr\,\sqrt{-g}\, T_{00}
\nonumber\\
&=& 4\pi\,v^2\int_0^\infty dr\, r^2\left[
\left(\frac{df}{dr}\right)^2 +
\frac{1}{r^2}\, f^2\,h^2
+ \frac{\lambda}{2}\,v^2\left(f^2 -1\right)^2\right]
\,\,\,\,\,\,.
\eea
The equation of motion (\ref{sep-eom-2}) together with $f(r)=\tanh(\alpha r)$ yield
\bea
E &=& 4\pi\,v^2\int_0^\infty dr\, \left[
 \frac{1}{2}  \dd_r\left(r^2\dd_r\,f^2\right)
-\frac{\lambda}{2}\,v^2\,r^2\left(f^4 -1\right)\right]
\,\,\,\,\,\,
\nonumber\\
&=&-4\pi\,v^2\alpha^2\int_0^\infty dr\,r^2 \left[\tanh^4\left(\alpha r\right) -1 \right]
\,\,\,\,\,\,.
\eea
The evaluation of the integral in the expression of $E$ involves 
the dilogarithm function ${\text{Li}}_2\left(z\right)$ and the final result is 
\be
E=\frac{4\pi}{3}\,\frac{v^2}{\alpha}\left(1+\frac{\pi^2}{3}\right) 
\,\,\,\,\,\,.
\ee
In reaching this result we have used the identity (see \cite{zagier} for instance)
\be
{\text{Li}}_2\left(z\right)=- {\text{Li}}_2\left(\frac{1}{z}\right) - \frac{1}{2}\,\ln^2\left(-z\right) 
- \frac{\pi^2}{6} \,\,\,\,\,\,
\ee
together with the special values ${\text{Li}}_2\left(0\right)=0$ and ${\text{Li}}_2\left(-1\right)=-\pi^2/12$.

In conclusion, we have found simple static solutions to the equations of motion of a complex scalar
field moving in a gauge field background. These are given in both cylindrical and spherical coordinates.
In cylindrical coordinates, the gauge field background could be interpreted as representing a vortex 
located in the $(r-\theta)$ plane. This configuration of fields  has a finite energy per unit length. 
Similarly, in spherical coordinates, the magnetic field is in the form of localised lumps in the
$(r-\theta)$ and $(\theta-\varphi)$ planes. In this case, the energy available in the whole space
is finite.

\section{Discussion}

There are various questions raised by the solutions presented in this paper\footnote{I am very grateful 
to the two anonymous reviewers for their scrutiny of the results of this note.}. We will discuss 
the settings in cylindrical coordinates as this is the one relevant to the Nielsen-Olesen vortices.
Let us recall briefly the situation when the gauge field $A_\mu$ is dynamical. In this case
the theory is the full  Abelian Higgs model as given by  the action 
\bea
S
= 
\int d^4x\sqrt{-g}\left[-\frac{1}{4}F_{\mu\nu}F^{\mu\nu}+
\left(D_\mu\phi\right)^\star \left(D^\mu\phi\right)
 -\frac{\lambda}{2}\left(\phi^\star \phi -v^2\right)^2\right]
 \,\,\,,
\label{A-H-M-1}
\eea
where $F_{\mu\nu} =  \dd_\mu A_\nu - \dd_\nu A_\mu$
is the field strength of the gauge field
$A_\mu $. 
\par
Using cylindrical coordinates together with 
the Nielsen-Olesen  ans\"{a}tze  (\ref{ansatz-A}), 
the equations of motion of the Abelian Higgs model become
\bea
{r}\,\frac{d}{dr}\left(\frac{1}{r}\frac{dh}{dr}\right) +2e^2v^2f^2\left(1-h\right) &=0&
\,\,\,\,\,\,,
\label{h-eq-2}
\\
\frac{1}{r}\,\frac{d}{dr}\left(r\frac{df}{dr}\right) -
n^2\,\frac{f}{r^2}\left(1-h\right)^2
 -\lambda v^2f\left(f^2-1\right) &=0&
\,\,\,\,\,\,.
\label{f-eq-2}
\eea
Similarly, the energy per unit length (along the  $z$ direction) of the Nielsen-Olesen vortex is 
\bea
E &=& 2\pi\int_0^\infty dr\, r\left[\frac{1}{r^2}\left( \frac{1}{2}\, \frac{n^2}{e^2}\,\left(\frac{dh}{dr}\right)^2 
+  n^2  v^2 f^2\,\left(1-h\right)^2\right)
+ v^2\left(\frac{df}{dr}\right)^2
+\frac{\lambda}{2}\,v^4\left(f^2 -1\right)^2\right]
\nonumber\\
&=& 2\pi\int_0^\infty dr\,\left\{
\frac{1}{4}\, \frac{n^2}{e^2}\dd_r\left[\frac{1}{r}\dd_r\left(\left(1-h\right)^2\right)  \right]
+r\left[  v^2\left(\frac{df}{dr}\right)^2
+\frac{\lambda}{2}\,v^4\left(f^2 -1\right)^2 \right]\right\}
\,\,\,\,\,\,.
\label{E-2}
\eea
The above equations correspond to the minimization of $E$ with respect to both  $h\left(r\right)$
and  $f\left(r\right)$. The second equality is a result of the use of the equations of motion.
\par
Since in our study the gauge field $A_\mu$ is taken to be a non-dynamical background, 
equation  (\ref{h-eq-2}) was not taken into account. This raises the question of how 
close the expressions
\bea
f\left(r\right) &=& \tanh\left(\alpha\,r\right)\,\,\,\,,
\nonumber \\
h\left(r\right) &=&  1\pm\,\frac{1}{n}\sqrt{\frac{2\alpha\,r}{\sinh\left(2\alpha\,r\right) }}
\,\,\,\,\,,\,\,\,\,\,\, \alpha^2=\frac{\lambda v^2}{2}
\,\,\,\,\,\,
\label{f-h}
\eea
are to the solutions of the equations of motion of the Abelian Higgs model ?
\par
Notice that for $n$ sufficiently large, the function $h\left(r\right)$ is nearly equal to  $1$.
Indeed, $h\left(r\right)$ is a monotonic function and varies between $1\pm\frac{1}{n}$ and $1$
for $r\in \left[0\,,\,\infty\right]$.
On the other hand, $h\left(r\right)=1$ is a solution to (\ref{h-eq-2}). Therefore, as a first interpretation, we could 
think of the two functions  $f\left(r\right)$ and $h\left(r\right)$ as approximate solutions
to the equations of motion of the Abelian Higgs model for large winding number $n$.
\par
The other feature of the study carried out in this article is that the energy 
in (\ref{E-cylinder}) is independent of the coupling constant $\lambda$.
This leads one to think that the system is in a critical regime very much like 
when the Bogomolny bound is saturated at the critical coupling $\lambda=e^2$
and where the  energy is $E=2\pi|n|v^2$ \cite{bogomolny}. It is tempting to speculate that
the critical regime in our analyses might correspond to  the formation of an aggregate
of a large number of vortices resulting in a single giant vortex. 
\par
Clusters of vortices, 
which are metastable, have been experimentally observed in condensed matter physics.
One of the early giant vortices was observed in the superfluid Helium $^4$He
with $n\approx 400$ \cite{marston-fairbank}.  More recently, dense arrangements 
of single Abrikosov vortices were detected in strongly confined superconducting
condensates \cite{cren-al}. Giant vortices (with $n$ ranging from $7$ up to $60$) were shown
to form in a rapidly rotating dilute-gas Bose-Einstein condensate \cite{engels-al}.  
It is well-known that the Abelian Higgs model (Ginzburg-Landau theory) provides the theoretical
framework for these condensed matter physics phenomena. Therefore, the results presented 
here could be of use as far as the formation of giant vortices is concerned.

\par
Another point regarding the solution presented in this note is that it depends very much on the 
expressions between the two brackets in (\ref{sep-eom}). In other words, different
assemblages of the terms are possible.
The only thing that one can say about this issue is  that equation (\ref{sep-eom}) gives 
the general expression of the gauge field background $h(r)$ in terms
of the dynamical field  $f(r)$. Indeed, we have 
\bea
\left(1-h\right)^2 = \frac{r^2}{n^2f^2}\left[ \frac{1}{2}
\frac{1}{r}\,\frac{d}{dr}\left(r\frac{df^2}{dr}\right) 
- \left(\frac{df}{dr}\right)^2 
 -\lambda v^2f^2\left(f^2-1\right)\right] 
\,\,\,\,\,\,.
\label{(1-h)}
\eea
The restrictions on the function  $f(r)$ are then : $i)$
the expression between brackets in  (\ref{(1-h)}) is always positive and 
 $ii)$ the energy 
\bea
E &=& 2\pi\,v^2\int_0^\infty dr\, \left[
 \frac{1}{2}  \dd_r\left(r\dd_r\,f^2\right)
-\frac{\lambda}{2}\,v^2\,r\left(f^4 -1\right)\right]
\eea
is finite and positive. 
The choice 
of the function  $f(r)$ (satisfying the points $i)$ and $ii)$ mentioned above) is
at the moment a matter of trial and error (and this is how we find our solution).

\par
One might also want to compare the expressions (\ref{f-h})  to the numerical
solutions of the Abelian Higgs model. Our function $f(r)$ and the magnetic field $B_r$ have  profiles similar to the one  found 
using numerical analyses, as can be seen from their sketshes 
in figure (\ref{f-B}). 
\par
Furthermore, we have injected the expressions of  $f(r)$ and $h(r)$ given in (\ref{f-h}) into
the gauge field equation of motion  (\ref{h-eq-2}) and evaluated this graphically. The equation of motion of the gauge
field is 
\bea
&\,&  \partial_\mu \left(\sqrt{-g}\, F^{\mu\nu}\right) +ie\sqrt{-g}\left[\phi\left(D^\nu\phi\right)^\star - 
\phi^\star\left(D^\nu\phi\right)\right] 
\nonumber \\
&=& -\frac{n}{e}\frac{1}{r} \left[r\frac{d}{dr}\left(\frac{1}{r}\frac{dh}{dr}\right) 
+2e^2v^2 f^2\left(1-h\right) \right] = 0 \,\,\,\,\,.
\label{h-eq-3}
\eea
Since $\left(1-h\right)$ contains a factor of $\frac{1}{n}$, the dependence on
the winding number cancels and the statement made below is independent of the value of $n$.
\par
\begin{figure}[thp]
\begin{center}
\includegraphics[scale=0.4]{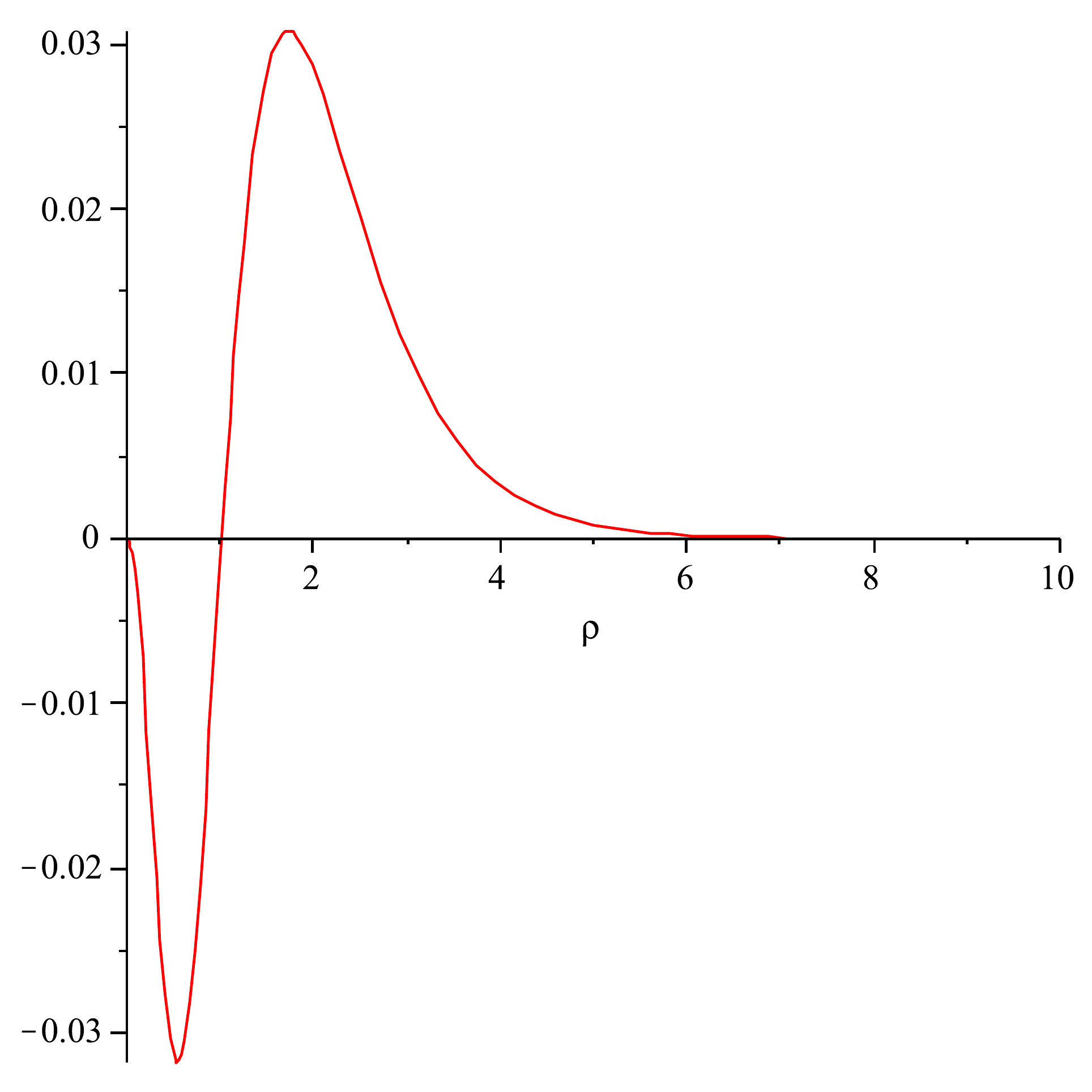}
\end{center}
\caption[]{The gauge field equation (\ref{h-eq-2}) for $\xi=\frac{\lambda}{e^2} =4.688$.
The horizontal axis is  the rescaled
coordinate $\rho = ev r$. }
\label{gauge-eq-1}
\end{figure}

We have first traded $r$ for the dimensionless coordinate $\rho = evr$. As a consequence, the only physical
coupling constant left in (\ref{h-eq-2}) and  (\ref{f-eq-2}) is $\xi=\frac{\lambda}{e^2}$.
We then tuned $\xi$ in such a way that the graph corresponding to the left hand side 
of  (\ref{h-eq-2}) is close to zero. Figure (\ref{gauge-eq-1}) is obtained for 
 $\xi=4.688$.
Therefore, one could affirm that the expressions of $f(r)$ and $h(r)$ in (\ref{f-h}) 
provide a relatively  good approximate solution to the equations of motion of the Abelian Higgs model for 
the particular value of the coupling constant  $\xi=4.688$.
\par
Using (\ref{f-h}) in  (\ref{E-2}), 
the corresponding energy is given by  
\bea
E &=&  \frac{2\pi}{3}\,{v^2}\left(\frac{\lambda}{2e^2} \right)  + \frac{2\pi}{3}\,{v^2}\left[-1 +4\ln\left(2\right)\right] 
\nonumber \\
&=& \frac{2\pi}{3}\,{v^2}\left[1.344 +4\ln\left(2\right)\right] 
\,\,\,\,\,\,.
\label{E-3}
\eea
This is to be compared to  $E=\frac{2\pi}{3}\,{v^2}\left[0.5 +4\ln\left(2\right)\right]$ as read from (\ref{E-cylinder}).
\par
In summary, the expressions of $f(r)$ and $h(r)$ given in (\ref{f-h}) furnish a quite good approximate 
solutions to the equations of motion of the Abelian Higgs model either for a large winding number
or for a special value of the coupling constant $\xi=\frac{\lambda}{e^2}$.
\par
Finally, we should mention that  the extension of the study carried out in this paper 
to non-Abelian gauge theories is currently under investigation. 
The prototype of such theories is the $SU(2)$ Georgi-Glashow \cite{georgi-glashow} model with a  
't Hooft-Polyakov ansatz \cite{thooft,polyakov} for the gauge field $A_\mu^a$ and the scalar field  $\phi^a$ ($a=1,2,3$).
If the gauge fields are fixed backgrounds then one has only one equation of motion to solve.
The whole issue is then to obtain a finite energy.

\end{document}